\documentclass[journal]{IEEEtran}
\usepackage{times,amsfonts,bbm,dsfont,amsmath,color,amssymb,graphicx,epsfig,multirow,float,algorithm,algorithmic,bm,cite,setspace}
\usepackage[normalem]{ulem}
\usepackage{makecell}
\usepackage{stfloats}
\usepackage{arydshln}
\usepackage{psfrag}
\usepackage{epstopdf}
\usepackage{stfloats}
\usepackage{color,cases}
\usepackage{subfigure}
\begin{document}
\title{Spectrum Sharing for Internet of Things: A Survey }

\author{\authorblockN{Lin Zhang, Ying-Chang Liang, and Ming Xiao}
\thanks{Lin Zhang and Ying-Chang Liang are with the Center for Intelligent Networking and Communications, University of Electronic Science and Technology of China, Chengdu, China, emails: linzhang1913@gmail.com and liangyc@ieee.org; Ming Xiao is with the School of Electrical Engineering, Royal Institute of Technology, KTH, email: mingx@kth.se.}
}\maketitle

\thispagestyle{empty}

\begin{abstract}
The Internet of Things (IoT) is a promising paradigm to accommodate massive device connections in 5G and beyond. To pave the way for future IoT, the spectrum should be planed in advance. Spectrum sharing is a preferable solution for IoT due to the scarcity of available spectrum resource. In particular, mobile operators are inclined to exploit the existing standards and infrastructures of current cellular networks and deploy IoT within licensed cellular spectrum. Yet, proprietary companies prefer to deploy IoT within unlicensed spectrum to avoid any licence fee. In this paper, we provide a survey on prevalent IoT technologies deployed within licensed cellular spectrum and unlicensed spectrum. Notably, emphasis will be on the spectrum sharing solutions including the shared spectrum, interference model, and interference management. To this end, we discuss both advantages and disadvantages of different IoT technologies. Finally, we identify challenges for future IoT and suggest potential research directions.
\end{abstract}

%\begin{keywords}
%Cellular spectrum, IoT, spectrum sharing, unlicensed spectrum.
%\end{keywords}

%\pagestyle{empty}
%\newpage

\section{Introduction}
% and collect useful data. With these data, it is possible to automatically optimize the systems through centralized data processing.

Recent years witnessed the emergence of various smart systems, such as smart home, smart healthcare, smart transportation, smart logistics, smart agriculture, and so on. To embrace the promise of these smart systems, it is critical to link \emph{end-devices} (EDs) in each system to the core Internet for centralized data-processing and control. This is significantly meaningful for both individual and environmental benefits. To achieve this goal, the Internet of Things (IoT) is extensively studied in both academia and industry \cite{IoT}. In particular, the IoT aims to connect everything to the core Internet in a wireless manner and is expected to be a dominant paradigm in 5G and beyond. A typical application is \emph{Unmanned Aerial Vehicle} (UAV), which collects data from sensors on the ground in a wireless manner \cite{UAV_1} \cite{UAV_2}. Notably, it is foreseen by the Europe Union that billions of EDs will be connected to the IoT by 2020 \cite{Spectrum_sharing}.

%However, the available spectrum is limited and it is impractical to allocate such amount of dedicated spectrum for IoT connections. Nevertheless, it is observed that the current spectrum resource is usually under-utilized. By performing the spectrum sharing between the IoT and existing networks,

Clearly, the key enabler of wireless connections is the spectrum resource. To pave the way for the IoT in 5G and beyond, the spectrum for IoT usages should be planed in advance. According to the study in \cite{Spectrum_sharing}, around 76 GHz spectrum resource is needed to accommodate massive IoT connections for an exclusive spectrum use. Nevertheless, this amount can be sharply reduced to 19 GHz by spectrum sharing. Thus, spectrum sharing is an preferable approach to cope with the conflicts between massive IoT connections and limited spectrum resource.

\begin{figure}[!tp]
            \centering
            \includegraphics[scale=0.55]{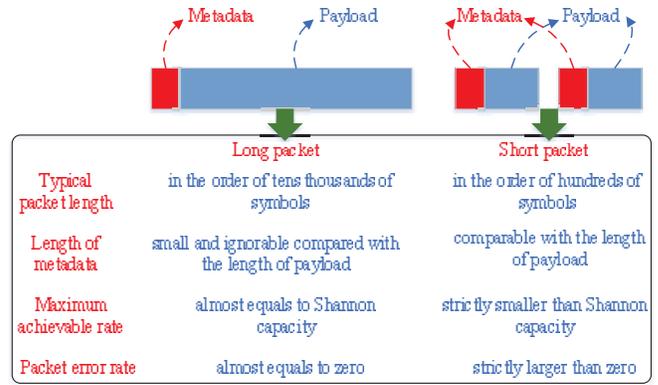}
            \caption{Comparison of the main features between long-packet and short-packet communications.}
            \label{Packet_length_comparison}
\end{figure}

%\begin{table}
%\label{Table_3}
% \caption{Comparison of long-packet and short-packet communications.}
% \centering
% \begin{tabular}{|p{2cm}|p{2.5cm}|p{2.5cm}|}
%%\begin{tabular}{|p{2cm}|p{7cm}|p{5cm}|}
%  \hline
%  % after \\: \hline or \cline{col1-col2} \cline{col3-col4} ...
%  Packet length & Long & Short \\
%    \hline
%  Typical length & in the order of tens thousands of symbols  & in the order of hundreds of symbols \\
%  \hline
%  Length of metadata & small and ignorable compared with the length of payload & comparable with the length of payload \\
%  \hline
%  Maximum achievable rate & almost equals to Shannon capacity & strictly smaller than Shannon capacity \\
%  \hline
%  Packet error rate &  almost equals to zero  & strictly larger than zero \\
%    \hline
%\end{tabular}
%\end{table}

In fact, IoT spectrum sharing is different from conventional spectrum sharing. Roughly, the differences are two-folds. The first one comes from the traffic model: conventional spectrum sharing technologies are mainly developed for the downlink long-packet communication from an \emph{access point} (AP) to mobile devices. However, IoT spectrum sharing will be dominated by the uplink short-packet communication from EDs to the AP. A comparison of the main features between long-packet and short-packet communications is illustrated in Fig. \ref{Packet_length_comparison}. From the figure, it is inappropriate/inefficient to directly apply the designs for the long-packet communication into the short-packet communication. In other words, the designs for conventional spectrum sharing are quite suboptimal for IoT spectrum sharing. The second difference lies in the device class: mobile devices in conventional spectrum sharing are normally expensive and equipped with strong signal-processing capabilities and re-chargeable batteries. However, the EDs in IoT spectrum sharing are expected to be cheap and thus usually have limited signal-processing capabilities and un-rechargeable batteries. This difference requires simple and efficient designs for IoT spectrum sharing to reduce both hardware costs and energy consumptions at EDs.

%In particular, typical short-range scenarios include personal area network (e.g., wearable devices) and local area network (e.g., smart homes). Typical long-range scenarios include the smart logistics, smart agriculture, and smart healthcare.

In reality, mobile operators are inclined to exploit the existing standards and infrastructures of current cellular networks and deploy IoT within licensed cellular spectrum. Proprietary companies however prefer to deploy IoT within unlicensed spectrum to avoid any licence fee. In this paper, we provide a survey of prevalent IoT technologies. In particular, we first introduce two IoT technologies deployed within cellular spectrum, i.e., \emph{enhanced machine type communication} (eMTC) and \emph{narrow-band} IoT (NB-IoT). Then, we elaborate four IoT technologies deployed within unlicensed spectrum including two conventional IoT technologies (i.e., Bluetooth and Zigbee) and two new IoT technologies (i.e., LoRaWAN and SigFox). To this end, we show an emerging IoT technology called ambient backscatter communication \cite{BC_Liu_2013}, which is able to exploit either licensed or unlicensed spectrum. Different from the existing literature, we put an emphasis on their spectrum sharing solutions including the shared spectrum, interference model, and interference management. Besides, we discuss both advantages and disadvantages of these IoT technologies. Finally, we identify challenges for future IoT and suggest potential research directions.

%
%According to the coverage, the IoT spectrum sharing scenarios can be roughly divided into two categories: Short-range scenarios in the order of meters or tens of meters and long range scenarios in the order of hundreds of meters and beyond. In this paper, we provide a survey of existing technologies for potential IoT usages. In particular, we first elaborate two conventional short-range technologies (Bluetooth and Zigbee) and an emerging short-range technique (backscattering communications \cite{BC_Liu_2013}). Then, we introduce a series of \emph{low-power wide-area} (LWPA) technologies \cite{LPWA_survey} including \emph{enhanced machine type communication} (eMTC), \emph{narrow-band} IoT (NB-IoT), LORAWAN, and SigFox. Notably, eMTC and NB-IoT are supposed to share the cellular spectrum bands by exploiting the existing standards and infrastructure of current cellular networks. Bluetooth, Zigbee, LORAWAN, and SigFox are deployed within unlicensed spectrum bands to avoid any payment for the license. Different from the existing literature, we put an emphasis on the spectrum sharing solution of each technology and analyze the adopted interference management schemes.

\section{IoT Technologies deployed within Licensed Cellular Spectrum}
Both eMTC and NB-IoT are representative IoT technologies deployed within licensed cellular spectrum (\cite{Primer_NB-IoT,IEEE_network}, and the references therein). In particular, eMTC and NB-IoT are standardized in \emph{long term evolution} (LTE) Release 13/14 by the \emph{third generation partnership project} (3GPP), which establishes IoT by exploiting the standardized solutions in cellular networks and the existing infrastructures as much as possible. Essentially, both eMTC and NB-IoT are designed to provide low hardware costs, low energy consumptions, wide coverages, and massive connections.

%\footnote{Recently, new features and enhanced functionalities (e.g., accurate positioning and multicast) are introduced to eMTC and NB-IoT in LTE Release 14. These new features and enhanced functionalities broaden both eMTC and NB-IoT use cases and improve the flexibly and efficiency of both protocols.}.

%In particular, the ED cost is expected to be 75-80 percent lower than that in \emph{general packet radio service} (GPRS) and \emph{global system for mobile communications} (GSM), the lifetime of an ED is around 10 years with 5 Watt Hour battery, 15$-$20 dB \emph{maximum coupling loss} (MCL) enhancement is supported with respect to regular cellular services, and around $1\times 10^4$ EDs/$\text{km}^2$ can be accommodated.

\begin{table}
\label{Table_1}
 \caption{Comparison between eMTC and NB-IoT.}
 \centering
 \begin{tabular}{|p{1.5cm}|p{2cm}|p{3.7cm}|}
%\begin{tabular}{|p{3cm}|p{5cm}|p{7cm}|}
  \hline
  Specification & eMTC &  NB-IoT\\
  \hline
  Bandwidth & 1.08 MHz & 180 kHz \\
  \hline
  Typical UE output power &  23 dBm, 20 dBm & 20 dBm, 14 dBm\\
    \hline
  Coverage  &  MCL is around 155 dB & MCL is around 164 dB (for standalone operation) \\
  \hline
  Typical modulation  & 16 QAM & BPSK, QPSK\\
  \hline
   Downlink  & OFDMA (15 KHz tone spacing)& OFDMA (15 KHz tone spacing)\\
  \hline
     Uplink  & SC-FDMA (15 kHz spacing) & single tone (3.75 kHz spacing or 15 kHz spacing); SC-FDMA (15 kHz spacing)\\
  \hline
       Power saving & PSM, DRX & PSM, DRX\\
  \hline
         Duplexing  & half-duplex with a single antenna & half-duplex with a single antenna\\
    \hline
         Date rate  & around 1 Mbps (uplink and downlink) & around 250 kbps (multi-tone, uplink and downlink), around 20 kbps (single-tone, uplink)\\
  \hline
           Spectrum sharing solution & LTE spectrum, centralized, in-band & LTE spectrum and re-farmed cellular spectrum, centralized, stand-alone/in-band/guard-band\\
  \hline

\end{tabular}
\end{table}

\subsection{Basic features of eMTC and NB-IoT}
Both eMTC and NB-IoT inherit the LTE system to a large extent, including numerologies, channel coding, rate matching, interleaving, and so on. This inheritance may accelerate the roll-out of technical specifications, the development and deployment of IoT products. Meanwhile, both eMTC and NB-IoT protocols simplify the LTE protocol to conform to the new traffic models and requirements in IoT. A brief comparison of both technologies is provided in Table I\footnote{In the table, MCL is short for maximum coupling loss, QAM is short for quadrature amplitude modulation, BPSK is short for binary phase shift keying, QPSK is short for quadrature phase shift keying, OFDMA is short for orthogonal frequency division multiple access, SC-FDMA is short for single carrier frequency division multiple access, PSM is short for power saving mode, and DRX is short for discontinuous reception.}. In the following, we describe basic features of both technologies with an emphasis on their spectrum sharing solutions.

\subsubsection{Shared features between eMTC and NB-IoT} Since both eMTC and NB-IoT are developed based on the LTE system, they share some common features. One distinctive feature is that, both protocols use OFDMA in the downlink and SC-FDMA in the uplink. To reduce hardware costs at EDs, both protocols simplify the functionalities by only supporting single-antenna and half-duplex operations. Meanwhile, both protocols adopt a narrow bandwidth, which reduces the complexities of both \emph{analog-to-digital conversion} (ADC) and \emph{digital-to-analog conversion} (DAC) at EDs. Note that, the main energy consumption of an ED in eMTC and NB-IoT is periodically listening to paging messages and performing link quality measurements, and the energy consumption for data transmissions is only a small fraction of the total consumed energy. Thus, to prolong the battery lifetime, both protocols adopt PSM and DRX. In particular, PSM keeps an ED attached to the network, but allows the ED to turn off all functionalities of paging listening and link quality measurements for energy saving. DRX allows the ED to negotiate with the network on the data reception phases, in which the ED can turn off the receiving functionality for energy saving. In addition, both protocols allow repeat transmissions for latency-tolerant EDs to extend the network coverage.
%
% although they can provide different \emph{maximum coupling loss} (MCL) enhancements (15 dB by eMTC and 20 dB by NB-IoT).

\subsubsection{Different features between eMTC and NB-IoT}: A major difference between eMTC and NB-IoT is the target spectrum: eMTC exploits the LTE spectrum, and NB-IoT exploits both the LTE spectrum and other re-farmed cellular spectrum. In particular, the bandwidths of eMTC and NB-IoT are 1.08 MHz and 180 kHz, respectively. Note that, the bandwidth of each GSM carrier is 200 kHz and the bandwidth of each physical resource block in LTE is 180 kHz. Thus, the bandwidth of eMTC spans six physical resource blocks of LTE in the frequency domain, meanwhile the bandwidth of NB-IoT enables it to perfectly coexist with both GSM and LTE. Another difference lies in the uplink transmission. In particular, only multi-tone SC-FDMA (with 15 KHz tone spacing) is supported in eMTC, while both single-tone (with 3.75 kHz or 15 kHz spacing) and multi-tone SC-FDMA (with 15 kHz spacing) are used for NB-IoT. This difference enables NB-IoT to use the spectrum more efficiently and schedule more EDs with a reduced data rate compared with eMTC.

%Besides, both mobility and voice LTE, namely, Vo-LTE, are supported in eMTC but not in NB-IoT in LTE Release 13.

%For instance, accurate positioning is provided for both eMTC and NB-IoT use cases that may benefit from the location information of EDs. Multicast is proposed in both eMTC and NB-IoT to complete the data transmission from one point to multiple points within one transmission period, thereby improving both energy and spectral efficiencies.
%some other enhanced functionalities (i.e., VoLTE enhancements and mobility enhancements) are suggested in eMTC. Also,

\begin{figure}[!tp]
            \centering
            \includegraphics[scale=0.45]{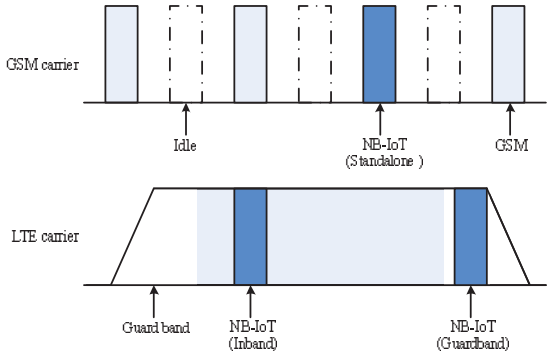}
            \caption{An example of three deployment modes.}
            \label{NB-IoT_spectrum_sharing}
\end{figure}

\subsection{Spectrum sharing solutions}
As aforementioned, eMTC aims to share the LTE spectrum, and NB-IoT aims to share both the LTE spectrum and other re-farmed cellular spectrum. This difference leads to distinct deployment modes of eMTC and NB-IoT. In particular, eMTC can only be deployed within the LTE spectrum, while NB-IoT has three deployment modes: stand-alone operation, in-band operation, and guard-band operation. In the stand-alone operation, NB-IoT is expected to use the idle cellular spectrum re-farmed from current cellular systems, e.g., GSM/CDMA. In the in-band operation, NB-IoT is deployed within LTE carriers and uses the PRBs same as LTE. In the guard-band operation, NB-IoT is deployed within the guard bands of LTE carriers. This is feasible since around $10\%$ bandwidth of a LTE carrier ($5\%$ on each side) is usually reserved to avoid the inter-carrier interference. An example of three deployment modes is provided in Fig. \ref{NB-IoT_spectrum_sharing}, where NB-IoT shares spectrum with GSM and LTE. It should be noted that, since both eMTC and NB-IoT share the spectrum with current cellular networks, mobile operators can manage the spectrum in a centralized manner. By allocating orthogonal spectrum to cellular/IoT links, mobile operators are able to avoid the interference between current cellular networks and eMTC/NB-IoT networks.

\section{IoT technologies deployed within unlicensed spectrum}

In this part, we elaborate four typical IoT technologies deployed within unlicensed spectrum, including two conventional IoT technologies (Bluetooth and Zigbee) and two new IoT technologies (LoRaWAN and SigFox).

\subsection{Bluetooth and Zigbee technologies}
Both Bluetooth and Zigbee are able to provide low hardware costs and low energy consumptions at EDs. As a result, Bluetooth and ZigBee have been widely applied in practical scenarios including wireless controls and wireless sensor networks. In the following, we focus on the spectrum sharing aspects of both technologies.

\subsubsection{Spectrum sharing solution of Bluetooth}
Bluetooth works in the \emph{Industrial Scientific Medical} (ISM) 2.4 GHz band (from 2.4-2.4835 GHz). Since the ISM 2.4 GHz band is globally unlicensed and free to access, Bluetooth is globally compatible and meanwhile suffers from a severe interference issue. To facilitate the spectrum sharing on the ISM 2.4 GHz band, Bluetooth adopts a \emph{frequency-hopping spread spectrum} (FHSS) approach. In particular, the FHSS is able to leverage frequency diversities and achieve spectrum sharing without spectrum planning, and thus is widely used to share unlicensed spectrum. For example, in the classic Bluetooth protocol, the spectrum between 2.4 GHz and 2.4835 GHz is divided into 79 1-MHz Bluetooth channels. If we define a \emph{hopset} as the set of channels used for hopping, a hopset may be partial or the whole of 79 Bluetooth channels. Besides, the transmitted data is divided into multiple packets and each packet is transmitted on one of the 79 Bluetooth channels based on a pre-determined order. To avoid severe interference from a Bluetooth transmission to other transmissions, it is regulated that the occupancy of each channel shall not be longer than 0.4 seconds.

%\footnote{Recently, an evolution of the classic Bluetooth, called \emph{Bluetooth low energy} (BLE), is proposed to support short-packet transmissions with a low power consumption and satisfy the requirements of IoT. Although the BLE divides the spectrum band from 2.4-2.4835 GHz into 40 2-MHz channels, it adopts an interference management scheme similar to the classic Bluetooth technology. Thus, we only elaborate the spectrum sharing solutions of the classic Bluetooth.}

%
%Bluetooth adopts a spread spectrum method. In particular, a spread-spectrum signal has noise-like properties and may slightly increase the background noise to a narrow-band receiver. Thus, the co-channel interference between a Bluetooth transmission and a narrow-band transmission can be efficiently reduced to the order of background noise power. Also, a spread-spectrum signal is resilient to a narrow-band co-channel interference.

\begin{figure}[!tp]
            \centering
            \includegraphics[scale=0.45]{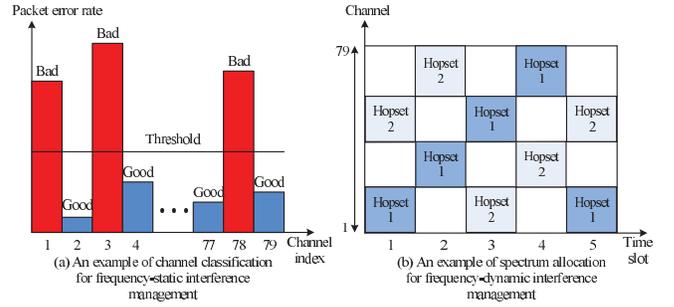}
            \caption{Examples to illustrate both the frequency-static and frequency-dynamic interference management schemes in Bluetooth communications.}
            \label{Bluetooth_interference_management}
\end{figure}

Nevertheless, Bluetooth devices may suffer from two kinds of interference: frequency-static interference and frequency-dynamic interference. In particular, the frequency-static interference occurs when a pair of Bluetooth transceiver hops to a channel, which is occupied by other transmissions. The frequency-dynamic interference comes from a collocated piconet, which is using identical channels. To combat the frequency-static interference, Bluetooth adopts an adaptive FHSS scheme, namely, AFHSS. The key idea of the AFHSS is as follows: the master node monitors the channel quality in an initial hopset during a certain time interval and classifies the channels in the initial hopset as ``good'' or ``bad''. By removing bad channels from the initial hopset, it is probable that only weak frequency-static interference exists on the remaining channels in the hopset. An example of the channel classification is shown in Fig. \ref{Bluetooth_interference_management}-(a), where the packet error rate together with a service-related threshold is used in the channel classification. To cope with the frequency-dynamic interference, a collaborative spectrum allocation scheme is proposed. Specifically, the time axis is first divided into orthogonal time slots with an identical duration. Then, the hopsets of multiple piconets are carefully designed in a collaborative manner, such that the channels in different hopsets are mutually orthogonal in each single time slot. Note that, the time synchronization among different piconets is required in this scheme. Otherwise, the asynchronization may lead to the non-orthogonality of the channels in different hopsets and inevitably results in frequency-dynamic interference. An example of the spectrum allocation for frequency-dynamic interference management is illustrated in Fig. \ref{Bluetooth_interference_management}-(b), where two piconets are considered and both hopsets are orthogonal in each time slot.

\subsubsection{Spectrum sharing solution of Zigbee}
Most commercial Zigbee devices share the ISM 2.4 GHz band and work on 16 2-MHz channels. To cope with the interference on the ISM 2.4 GHz band, Zigbee adopts a \emph{direct-sequence spread spectrum} (DSSS) technique and a \emph{carrier sense multiple access/collision avoidance} (CSMA/CA) access mechanism. In particular, by adopting the DSSS technique, Zigbee spreads a narrow band signal over a wide band channel with designed spread sequences. Then, the DSSS signal has noise-like properties and thus is resilient to the narrow band interference. Besides, the DSSS allows multiple EDs to access a common coordinator/router simultaneously. Although multiple signals may interfere with each other at a coordinator/router, the correlation property of spread sequences enables the coordinator/router to correctly extract the required signal from multiple signals. Furthermore, by using the CSMA/CA access mechanism, the Zigbee device first senses the target channel and precedes to the data transmission only if the target channel is sensed to be idle. In fact, the CSMA/CA is an effective channel access mechanism when the target channels are not crowd.

\subsection{LoRaWAN and SigFox technologies}
LoRaWAN and SigFox are respectively proposed by LoRa Alliance and SigFox company, both of which aim to deploy IoT within free ISM spectrum and reduce deployment costs. Due to the fact that the ISM spectrum in $2.4$ G and $5$ G have weak permeation capabilities, LoRaWAN and SigFox use the sub-GHz ISM bands to provide wide coverages. A comparison of both technologies is provided in Table II\footnote{In the table, MCL is short for maximum coupling loss, CSS is short for chirp spread spectrum, DBPSK is short for differential binary phase shift keying, GFSK is short for Gaussian frequency shift keying, PSM is short for power saving mode, and DRX is short for discontinuous reception.}. In the following, we describe the basic features of both technologies with an emphasis on their spectrum sharing solutions.

\begin{table}
\label{Table_2}
 \caption{Comparison between LoRaWAN and SigFox. }
 \centering
 \begin{tabular}{|p{1.3cm}|p{3.5cm}|p{2.5cm}|}
%\begin{tabular}{|p{2cm}|p{7cm}|p{5cm}|}
  \hline
  Specification & LoRaWAN &  SigFox\\
    \hline
  Band & sub-GHz: 433 MHz and 868 MHz in Europe, 915 MHz in the US, and 430 MHz in Asia & sub-GHz: 868 MHz in Europe and 902 MHz in the US \\
  \hline
  Bandwidth & 125 kHz & 100 Hz \\
  \hline
  Typical UE output power &  14 dBm & 20 dBm, 14 dBm\\
    \hline
  Coverage  &  MCL is 157 dB & MCL is 160 dB \\
  \hline
  Typical modulation  & CSS & DBPSK (uplink) and GFSK (downlink)\\
  \hline
       Power saving  & PSM, DRX & PSM, DRX\\
  \hline
         Date rate  & 0.3-37.5 kbps & 100 bps (uplink) and 600 bps (downlink) \\
  \hline
         Access mechanism  & ALOHA & ALOHA  \\
  \hline
           Spectrum sharing solutions  & duty cycle no larger than $1\%$ and reception diversity &  duty cycle no larger than $1\%$ \\
  \hline

\end{tabular}
\end{table}

\subsubsection{Basic features of LoRaWAN and SigFox}
To achieve low energy consumptions at EDs, LoRaWAN divides EDs into three categories based on the downlink latency requirement: Class A, Class B, and Class C. In particular, the EDs in Class A are insensitive to the downlink latency. These EDs open downlink receive windows only after a uplink transmission. In other words, most functionalities of these EDs can be switched off for energy saving if there is no uplink transmission. The EDs in Class B have limited requirements on the downlink latency. These EDs are scheduled to open downlink receive windows periodically. In fact, most functionalities of these EDs can still be switched off for energy saving if they are not scheduled meanwhile there is no uplink transmission. The EDs in Class C have stringent requirements on the downlink latency. These EDs shall always keep receive windows open and only close them for uplink transmissions. Besides, LoRaWAN adopts an ALOHA access mechanism to simplify the operations at EDs. That means, an ED transmits its data directly without checking the channel state and re-transmits the data if a collision happens. To improve the chance of correctly decoding the required signal at the AP, LoRaWAN adopts a retransmission mechanism with a maximum retransmission number of eight. To achieve a wide coverage, LoRaWAN adaptively changes the data rate. In particular, LoRaWAN has six classes of data rates ranging from 0.3 kbps and 37.5 kbps depending on different spreading factors and channel bandwidths. In particular, a small data rate corresponds to a wide coverage and vice versa.

Two highly recognizable features of SigFox are the ultra-narrow bandwidth, i.e., $100$ Hz, and the cognitive functionality at each AP. In particular, by using the ultra-narrow bandwidth, SigFox reduces the background noise power at both AP and ED. This saves ED's transmit power and extends AP's coverage with a MCL around 160 dB. Meanwhile, ultra-narrow bandwidth means a low requirement on the signal-processing capability and thus reduces the hardware cost (e.g., DAC and ADC) at each ED. However, the ultra-narrow bandwidth also leads to an achievable data rate as low as 100 bps. With the cognitive functionality, each AP is able to automatically identify the channel that an ED uses. Then, it is unnecessary to exchange channel selection signallings, thereby saving both power and spectrum resources. Similar to LoRaWAN, SigFox also adopts an ALOHA access mechanism for uplink transmissions. But, SigFox does not support the acknowledgement for each uplink transmission. In other words, an ED cannot be informed of a failure in the uplink transmission. To cope with the issue, SigFox enables an ED to transmit each signal for three times and improves the chance of correctly decoding the required signal at the AP.

\subsubsection{Spectrum sharing solutions of LoRaWAN and SigFox}
LoRaWAN works on various sub-GHz ISM bands in different countries/regions. In particular, LoRaWAN works on 433 MHz and 868 MHz in Europe, 915 MHz in the US, and 430 MHz in Asia. To comply with the spectrum regulation and avoid substantial interference to other electronic devices on the same spectrum/channel, LoRaWAN is allowed to work with a duty cycle no larger than $1\%$. Besides, LoRaWAN adopts a proprietary CSS technique and spreads a narrow band signal over a wideband channel. Then, the LoRaWAN is resilient to the narrow band interference. Furthermore, LoRaWAN has a star-of-star topology, in which each UE is connected to multiple APs and each AP is also connected to multiple EDs. In other words, a transmit signal from an ED may be received by multiple APs. By combining the received signals in multiple APs, LoRaWAN enhances the chance to successfully decode the signal and creates a reception diversity gain.

Similar to the LoRaWAN, SigFox also works on the sub-GHz ISM bands. In particular, SigFox works on 868 MHz in Europe and 902 MHz in the US. To comply with the spectrum regulation and avoid substantial interference to other electronic devices on the same spectrum/channel, SigFox is allowed to work with a duty cycle no larger than $1\%$. Notably, SigFox can transmit 140 12-byte messages per day in the uplink and 4 8-bytes messages per day in the downlink. Besides, SigFox enables an ED to transmit multiple copies of a signal on different channels and combines these copies at the AP in the presence of interference. This improves the chance to correctly decode the required signal at the AP and leverages a reception diversity gain.

\section{Ambient Backscatter communication}

In previous sections, we have introduced IoT technologies deployed in licensed and unlicensed spectrum, respectively. In this section, we will elaborate a promising IoT technology, i.e., ambient backscatter communication, which can be flexibly deployed in either licensed or unlicensed spectrum. In particular, ambient backscatter communication is an effective short-range transmission solution with both ultra-low hardware costs and ultra-low energy consumptions at EDs. These features make the ambient backscatter communication quite suitable to IoT. In the following, we first provide the basic principle of the ambient backscatter communication. Then, we analyze the spectrum sharing solutions of different ambient backscatter communication scenarios in the states of the arts.

%Meanwhile, by sharing the spectrum of ambient wireless signals (instead of the dedicated spectrum), ambient backscatter communication also improves the spectrum efficiency.

\subsection{Basic principle}

%Essentially, an ambient backscatter transmitter, namely, tag, is mainly consisted of a digital logic integrated circuit and an antenna, and differs from a conventional radio transmitter which usually has multiple RF components (such as oscillator, digital-analog converter, amplifier, filter, mixer). This reduces both the hardware cost and power consumption of each ED.

In general, ambient backscatter communication has two successive phases: energy acquisition and data transmission \cite{BC_Liu_2013}. In the energy acquisition phase, the transmitter, namely, tag, harvests energy from ambient wireless signals (i.e., electromagnetic waves) and activates the internal circuit. In the data transmission phase, the tag modulates data on ambient wireless signals. In fact, it is possible to modulate data on ambient wireless signals. Intuitively, when an electromagnetic wave travels from media A to media B, a partial electromagnetic wave may be reflected from the boundary of media B back to media A if two media have different impedances/densities. Thus, by adjusting the impedance of the tag antenna in the presence of an ambient electromagnetic wave, the intended receiver of the tag, namely, reader, may detect different reflected electromagnetic waves in terms of the amplitude and/or the phase, which can be used to map the transmitted/received data at the tag/reader.

Therefore, the ambient backscatter communication is able to realize the data transmission from a tag to a reader without complicated \emph{radio frequency} (RF) components at the tag. A typical tag is mainly consisted of a digital logic integrated circuit and an antenna. This sharply reduces both the hardware cost and energy consumption at the ED compared with a conventional radio transmitter. It should be noted that, the shared spectrum which conveys the ambient wireless signal could be either licensed or unlicensed spectrum.

%Nevertheless, since the reflected signal is weak, the communication distance between a tag and a reader is in the order of meters.

%Specifically, if the tag is close to the RF source, it is possible for the tag to capture energy from the ambient electromagnetic wave according to the electromagnetic theory. The maximum distance between the RF source and the tag is related to the frequency of the electromagnetic wave as well as the transmit power of the RF source. In particular, the maximum distance is large when the frequency is small or the transmit power is large.

\subsection{Spectrum sharing solutions}
%Since backscatter communication devices use the ambient electromagnetic wave for data transmission and do not need dedicated spectrum resource, the backscatter communication can be viewed as a spectrum sharing technology. The shared spectrum is determined by the ambient electromagnetic wave, and can be licensed cellular spectrum bands, TV bands, as well as unlicensed WiFi bands. In the following, we analyze the existing literature on the backscatter communication from the aspect of spectrum sharing.

\begin{figure}[!tp]
            \centering
            \includegraphics[scale=0.6]{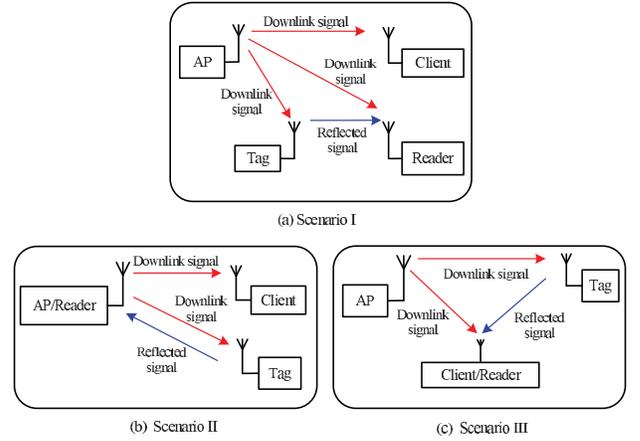}
            \caption{Three ambient backscattering communication scenarios: a) the AP is transmitting data to the client, and meanwhile acts as a RF source for the tag, which exploits the wireless signals from the AP and delivers data to the reader \cite{BC_Liu_2013} \cite{Yang_TCOM_2017} \cite{Kang_submitted}; b) the AP is transmitting data to the client, and meanwhile acts as a reader, which needs to receive data from the tag \cite{BackFi} \cite{FD_katti}; c) the AP is transmitting data to the client, which acts as a reader and needs to receive data from the tag \cite{Yang_submitted}.}
            \label{Backscattering_scenarios}
\end{figure}

%Scenario I consists of a reader and a tag. In particular, the reader also acts as the RF source and needs to send a wireless signal to initialize a backscatter communication. By capturing energy from the wireless signal, the tag is powered on and can modulate the information bits on the wireless signal.

According to various practical applications, the existing literature on the ambient backscatter communication can be divided into three scenarios as shown in Fig. \ref{Backscattering_scenarios}. Scenario I consists of four nodes: an AP, a client, a reader, and a tag. In particular, the AP is transmitting data to the client, and meanwhile acts as a RF source for the tag. The tag exploits the wireless signals from the AP and delivers data to the reader. Scenario II consists of three nodes: an AP, a client, and a tag. In particular, the AP is transmitting data to the client, and meanwhile acts as a reader, which needs to receive data from the tag. Scenario III consists of three nodes: an AP, a client, and a tag. In particular, the AP is transmitting data to the client, which acts as a reader and needs to receive data from the tag.

\subsubsection{Spectrum sharing solution in Scenario I} In this scenario, the reader can receive the downlink signal from the AP and the reflected signal from the tag simultaneously. In particular, the reader needs to detect the required data on the reflected signal. Clearly, the downlink signal from the AP may interfere with the detection of the required data on the reflected signal. According to whether or not to cancel the downlink signal from the received signal before detecting the required data, the existing literature can be divided into two categories. The first category treats the downlink signal as interference and detects the required data in the presence of the interference \cite{BC_Liu_2013}. The second category first cancels the interference and then detects the required data \cite{Yang_TCOM_2017} \cite{Kang_submitted}.

% Note that, Since the tag is close to the AP, the reader is also close to the wireless AP. Then, the direct interference dominates the received interference at the reader. Based on the adopted management schemes of the direct interference, the existing literature can be divided into two categories. The first category treats the direct interference as the background noise. The second category intends to cancel the direct interference.

Specifically, \cite{BC_Liu_2013} combines the on-off (reflect-unreflect) keying modulation and the differential coding at the tag. By measuring the variation of received power levels, the reader can extract the required data in the presence of the interference. Note that, the channel state information from the RF source and the reader is not needed at the reader. \cite{Yang_TCOM_2017} considers an ambient OFDM signal and exploits the repeated cyclic prefix in an OFDM symbol to cancel the interference of the OFDM signal. \cite{Kang_submitted} considers that the downlink signal is usually much stronger than the reflected signal, and adopts a successive interference cancellation method to accelerate data detections. Specifically, the reader first detects the downlink signal by treating the reflected signal as interference. Then, the reader detects the required data on the reflected signal after recovering the downlink signal and subtracting it from the received signal.

%and transmit bits according to the variation to transmit bits the tag modulates $1$ or $0$ on the ambient wireless signal by switching on or off the impedance. Since the power of the received signal with and without the reflected copy at the reader may be different, the reader can adopt an energy detector to decode the data bits on the reflected signal.

%However, the direct interference cancellation algorithm is only developed when the wireless access point transmits an OFDM signal \cite{Yang_TCOM_2017}. The direct interference cancellation algorithms for other signals are still open problems.

\subsubsection{Spectrum sharing solution in Scenario II} In this scenario, the AP works in a full-duplex mode, i.e., transmit data to the client and receive data from the tag simultaneously \cite{BackFi}. Clearly, when the AP detects the required data on the reflected signal, it suffers from the self-interference caused by the downlink signal, which constrains the achievable rate from the tag to the reader.

%Specifically, classical self-interference cancellation techniques are realized in both analog domain and digital domain. In the analog-circuit domain, the analogy form of the self-interference signal is first predicted and then subtracted from the received analog signal. Due to the imperfect processing of analog components, the self-interference cannot be completely cancelled in the analog domain. Then, the digital-domain cancellation is followed to cancel the residual self-interference. In particular, digital-domain cancellation means estimating the digital form of the residual self-interference after the analog-domain cancellation and subtracting the predicted signal from the baseband samples in digital domain.

%the  In fact, the self-interference can be cancelled in both analog domain and digital domain based on the classical self-interference cancellation techniques. Unfortunately,

Although extensive literature has studied the self-interference cancellation problem in full-duplex communications \cite{FD_katti}, classical self-interference cancellation techniques cannot be directly applied in this scenario. The reason is as follows: in conventional full-duplex communications, the required signal component is independent from the self-interference. Then, the required signal component remains after cancelling the self-interference from the received signal. In Scenario II, the required signal component is the reflected signal, which is highly correlated to the self-interference. Thus, classical self-interference cancellation techniques may cancel both self-interference and the required signal component, i.e., the reflected signal, degrading the detection performance of the required data at the AP. To cope with this issue, an optimized link layer design is provided to prevent the self-interference cancellation from cancelling the required signal component in \cite{BackFi}.

\subsubsection{Spectrum sharing solution in Scenario III} In this scenario, the client receives the downlink signal from the AP and the reflected signal from the tag simultaneously \cite{Yang_submitted}. Since two signals may interfere with each other, the optimal detector needs to detect both signals in a joint manner. To deal with the issue, \cite{Yang_submitted} proposes an optimal high-complexity \emph{maximum likelihood} (ML) detector. However, the computational complexity of the ML detector grows exponentially with the modulation size of the downlink signal and is hard to implement in practical situations. Alternatively, a sub-optimal low-complexity detector based on the successive interference-cancellation is also developed.

\begin{table}
\label{Table_2}
 \caption{Comparison of different IoT technologies.}
 \centering
 \begin{tabular}{|p{1.8cm}|p{1.8cm}|p{1.85cm}|p{1.85cm}|}
%\begin{tabular}{|p{2cm}|p{7cm}|p{5cm}|}
  \hline
  IoT technologies & Shared spectrum & Advantages &  Disadvantages\\
    \hline
  eMTC, NB-IoT & licensed cellular & wide coverage, QoS guarantee & expensive\\
  \hline
  Bluetooth, ZigBee, LoRaWAN, SigFox & unlicensed ISM & free licence, wide coverage & limited or no QoS guarantee \\
  \hline
  Ambient backscatter communication &  licensed or unlicensed spectrum  & ultra-low hardware cost, ultra-low energy consumption & short transmission distance \\
    \hline
\end{tabular}
\end{table}

\section{Comparison among different IoT technologies}

From the analysis in the previous parts, mobile operators are inclined to take advantage of the existing standards and infrastructures of current cellular networks and deploy IoT within cellular spectrum. Although it is expensive to share the cellular spectrum for IoT applications, it has two main advantages. Firstly, by reusing the current cellular resource, mobile operators can achieve a wide coverage of IoT in a short time. This accelerates the roll-out of IoT products. Secondly, the coexistence between IoT and cellular networks can be achieved in a centralized manner. This eases the interference management between IoT and cellular networks, and provides \emph{quality of service} (QoS) guarantees for both IoT and cellular links.

On the other hand, proprietary technologies prefer to use the unlicensed ISM spectrum to avoid any licence fee. However, the interference on unlicensed ISM bands are usually unpredictable and complicated. Thus, proprietary technologies need to adopt smart interference cancellation/mitigation approaches to improve the decoding performance at an IoT receiver. Still, existing proprietary technologies can provide limited QoS guarantees, or cannot provide any QoS guarantee.

%Thus, proprietary technologies avoid the licence cost at the expense of QoS guarantees.

Additionally, the ambient backscatter communication can share both the cellular spectrum and unlicensed ISM spectrum. Although the ambient backscatter communication can largely reduce the hardware cost and energy consumption at the tag, the interference model at the reader is quite complicated. Then, advanced interference management schemes shall be adopted at the reader. Meanwhile, the ambient backscatter communication exploits the reflected signal to transmit data and suffers from a short transmission distance (in the order of meters). To summarize, we provide a brief comparison of different IoT technologies in Table III.

\section{Other potential spectrum sharing technologies for future IoT}

Other potential spectrum sharing technologies for future IoT include \emph{cognitive radio} (CR), \emph{device-to-device} (D2D), \emph{non-orthogonal multiple access} (NOMA), LTE \emph{on unlicensed spectrum} (LTE-U), and so on \cite{Lin_2017}. Specifically, CR is a suitable solution in a hierarchial network in which primary users have priorities to use some certain spectrum while secondary users are allowed to access the spectrum in an opportunistic manner. D2D is proposed to directly connect neighbouring devices for data exchanges and differs from the conventional cellular network, in which devices are intermediately connected through a \emph{base station} (BS). NOMA is treated as a spectrum sharing technology due to the fact that it allows multiple devices to be scheduled on the same spectrum (channel) simultaneously. It is worth noting that, compared with conventional orthogonal multiple access technologies (e.g., time/frequency/code division multiple access), NOMA is able to provide a significant performance gain when simultaneously scheduling the devices with different types of applications/data rates. LTE-U is supposed to offload the data traffic on licensed spectrum by integrating the licensed spectrum with unlicensed one for data transmissions (please refer to \cite{LTE-U} and the references therein for more details).

%Therefore, the backscattering communication avoids the licence cost and reduces the energy consumption with a decreased data rate, a reduced transmission distance, and an increased interference management complexity.

%Consequently, different adopted spectrum together with different interference management schemes may lead to different balances between costs (e.g., spectrum license and circuit design) and QoS (e.g., data rate and latency).

\section{Challenges and Research Directions}

It is envisioned that the ever expanding IoT will generate a massive volume of data, which has diverse requirements including latency, reliability, and throughput. In this section, we start with a typical use case of future IoT, which cannot be well addressed by the existing IoT technologies. Then, we identify potential research directions/solutions. Finally, we list several other challenges and research directions.

\subsection{Use case analysis }

Suppose that a mobile operator (namely, MO-A) exploits existing BSs to first collect the data from registered IoT devices and then forward them to a server for further data-processing. In particular, some specific IoT devices, which are in the coverage of a BS belonging to MO-A, have stringent latency requirements on both uplink data collections and downlink responses. Due to the limited spectrum resource and the dynamic nature of the data traffic at the BS, the requirements of these IoT devices cannot be always satisfied. Apparently, the existing IoT technologies are not applicable in this case. Next, we identify two potential research directions/solutions to address this issue.

\subsection{Inter-operator spectrum sharing}
The first potential research directions/solution is leveraging the inter-operator spectrum sharing, i.e., spectrum sharing among different mobile operators. Specifically, if these IoT devices are simultaneously covered by the BS of MO-A and another BS, which is underutilized in terms of the spectrum resource but belongs to another mobile operator (namely, MO-B), it is possible to satisfy these IoT devices with the spectrum resource of MO-B. To achieve this, the spectrum sharing between MO-A and MO-B is required. In fact, the inter-operator spectrum sharing means a more flexible utilization of limited spectrum resource compared with the intra-operator spectrum sharing, i.e., spectrum sharing within a single mobile operator. Nevertheless, the interference of inter-operator spectrum sharing shall be carefully coordinated to guarantee the system performance of each operator.

\subsection{Edge computing}
An alternative research directions/solution is re-designing the architecture of the network for edge computing. Specifically, by deploying edge data-processing nodes close to IoT devices, the latency requirements of these IoT devices can be addressed from two aspects. Firstly, some data can be directly collected and responded by edge data-processing nodes. Secondly, some raw data from IoT devices can be pre-processed at edge data-precessing nodes and only the useful part will be forwarded to the BS, reducing the traffic at the BS. In this way, the BS, edge data-processing nodes, and IoT devices form a hierarchical IoT network, in which the interference is complicated and needs to be carefully managed.

%
%
%
%  through spectrum sharing. It is worth pointing out that, the existing spectrum sharing solutions are only applicable in a specific IoT network subject to a certain mobile operator or proprietary company, and cannot deal with the spectrum sharing issue among different IoT networks.
%
%
%
%In the previous sections, we introduce several promising IoT technologies with an emphasis their spectrum sharing solutions. Yet, each spectrum sharing solution is only applicable in a specific IoT network subject to a certain mobile operator or proprietary company, and cannot deal with the spectrum sharing issue among different IoT networks. Future IoT is envisioned to be heterogeneous and it is probable that some scenarios in future IoT need the spectrum sharing among different IoT networks, which belongs to different mobile operators or proprietary companies.

\subsection{Other Challenges and Research Directions}

%\begin{enumerate}
\subsubsection{More use cases} Future IoT must be heterogeneous and include new network topology/traffic model/interference model, which cannot be well settled with the current IoT technologies. For example, the design and analysis for real-time IoT are rare in the existing literature \cite{Real_time_IoT_survey}. Thus, one potential research direction is to explore more IoT use cases and rethink their spectrum sharing solutions.

\subsubsection{More theoretical analysis and design} IoT traffic is dominated by short-packet communications, which differs from conventional long-packet communications as shown in Fig. 1. It is inappropriate and inefficient to directly apply the theoretical analysis and design of conventional spectrum sharing into IoT scenarios. Thus, another research direction is to conduct more theoretical analysis and design of spectrum sharing for short-packet communications.

\subsubsection{More spectrum resource for IoT sharing} It is clear that, if the available spectrum resource is not enough, the spectrum sharing is not going anywhere. With the increasing demand of IoT traffic, it is urgent to explore more spectrum resource for IoT spectrum sharing and accommodate more IoT traffic. For example, Millimeter wave is an effective high-rate short-range communication technology and can be used in some IoT scenarios. Nevertheless, the spectrum sharing on Millimeter wave is challenging since its interference model is quite different from the existing IoT technologies. Thus, it is necessary to explore more spectrum resource for IoT usages and analyze the corresponding spectrum sharing solutions.

%\end{enumerate}

\section{Conclusions}

In this paper, we provided a survey on the existing IoT technologies including eMTC, NB-IoT, Bluetooth, ZigBee, LoRaWAN, SigFox, and ambient backscatter communication. In particular, eMTC and NB-IoT are deployed within licensed cellular spectrum. Bluetooth, ZigBee, LoRaWAN, and SigFox are deployed within unlicensed spectrum. Ambient backscatter communication can be deployed within either licensed or unlicensed spectrum. For each IoT technology, we first elaborated the basic features/principles and then analyzed the spectrum sharing solution including the shared spectrum, interference model, and interference management scheme. After that, we discussed both advantages and disadvantages of these spectrum sharing solutions. Finally, we identified challenges for future IoT and suggested potential research directions.

%\bibliographystyle{IEEEtran}
%\bibliography{IEEEabrv,Ref}

\begin{thebibliography}{1}

\bibitem{IoT} M. R. Palattella , M. Dohler, A. Grieco, G. Rizzo, J. Torsner, T. Engel, and L. Ladid, ``Internet of things in the 5G era: enablers, architecture, and business models,'' \emph{IEEE J. Sel. Areas Commun.}, vol. 34, no. 3, pp. 510-527, Mar. 2016.

\bibitem{UAV_1} F. Tang,  Z. M. Fadlullah, N. Kato, F. Ono, and R. Miura, ``AC-POCA: anticoordination game based partially overlapping channels assignment in combined UAV and D2D-based networks,'' \emph{IEEE Trans. Veh. Technol.}, vol. 67, no. 2, pp. 1672-1683, Feb. 2018.

\bibitem{UAV_2} D. Takaishi, Y. Kawamoto, H. Nishiyama, N. Kato, F. Ono, and R. Miura, ``Virtual cell-based resource allocation for efficient frequency utilization in unmanned aircraft systems,'' \emph{IEEE Trans. Veh. Technol.}, Vol. 67, no. 4, pp. 3495-3504, Apr. 2018.


\bibitem{Spectrum_sharing} Europe Union, Identification and quantification of key socio-economic data to support strategic
planning for the introduction of 5G in Europe - SMART 2014/0008, 2016.

%
%\bibitem{Cisco}Cisco, The Internet of Things: How the next evolution of the Internet is changing everything, 2011.

\bibitem{BC_Liu_2013} V. Liu, A. Parks, V. Talla, S. Gollakota, D. Wetherall, and J. R. Smith, ``Ambient backscatter: wireless communication out of thin air,'' in \emph{Proc. of ACM SIGCOMM}, Hong Kong, China, Jun. 2013, pp. 1-13.




\bibitem{Primer_NB-IoT} Y.-P. E. Wang et al., ``A primer on 3GPP narrowband Internet of Things (NB-IoT),'' \emph{IEEE Commun. Mag.}, vol. 55, no. 3, pp. 117-123,  Mar. 2017.

\bibitem{IEEE_network} A. Hoglund et al., ``Overview of 3GPP Release 14 enhanced NB-IoT,'' \emph{IEEE Network}, vol. 21, no. 6, pp. 16-22, Nov. 2017.

%\bibitem{Bluetooth_SP} The Bluetooth Special Interest Group, ``Specification of the Bluetooth
%system, covered core package,'' 2010.
%
%\bibitem{ZigBee_SP} ZigBee Alliance, ZigBee Specfication, Jan. 2008.
%
%\bibitem{LoRaWAN} N. Sornin, M. Luis, T. Eirich, T. Kramp, and O.Hersent, LoRaWAN Specfication, ver. 1.0, Jan. 2015.
%
%
%\bibitem{SigFox} SigFox, Accessed Dec. 10 2017. [Online]. Available: https://www.sigfox.com/.

%
%\bibitem{LPWA_survey} U. Raza, P. Kulkarni, and M. Sooriyabandara, ``Low power wide area networks: an overview,'' \emph{IEEE Commun. Surveys $\&$ Tutorials}, vol. 19, no. 2, pp. 1018-1044, Second quarter, 2017.


% \bibitem{RFID} K. Finkenzeller, ``RFID handbook, third edition,'' Wiley press.

\bibitem{Yang_TCOM_2017} G. Yang, Y.-C. Liang, R. Zhang, and Y. Pei, ``Modulation in the air: backscatter communication over ambient OFDM carrier,'' \emph{IEEE Trans. Commun.}, vol. 66, no. 3, pp. 1219-1233, Mar. 2018.

 \bibitem{Kang_submitted} X. Kang, Y.-C. Liang, and J. Yang, ``Riding on the primary: a new spectrum sharing paradigm for wireless-powered IoT devices,'' \emph{Proc. of IEEE ICC} 2017.


 \bibitem{BackFi} D. Bharadia, K. Joshi, M. Kotaru, and S. Katti, ``BackFi: high throughput WiFi backscatter,'' in \emph{Proc. of ACM SIGCOMM}, London, UK, Aug. 2015, pp. 283-296.

\bibitem{FD_katti} D. Bharadia, E. McMilin, and S. Katti, ``Full duplex radios,'' in \emph{Proc. ACM SIGCOMM}, Hong Kong, China, Aug. 2013, pp. 375-386.


 \bibitem{Yang_submitted} G. Yang, Q. Zhang, and Y.-C. Liang, ``Cooperative ambient backscatter communication systems for internet-of-things,'' \emph{IEEE Internet of Things}, vol. 5, no. 2, pp. 1116-1130, Apr. 2018.


\bibitem{Lin_2017} L. Zhang, M. Xiao, G. Wu, M. Alam, Y.-C. Liang, and S. Li, ``A survey of advanced techniques for spectrum sharing in 5G networks,'' \emph{IEEE Wireless Commun.}, vol. 24, no. 5, pp. 44-51, Oct. 2017.

 \bibitem{LTE-U} H. Zhang, Y. Liao, and L. Song, ``D2D-U: Device-to-fevice communications in unlicensed bands for 5G system,'' \emph{IEEE Trans. Wireless Commun.}, vol. 16, no. 6, pp. 3507-3519, Jun. 2017.

\bibitem{Real_time_IoT_survey} S. Verma, Y. Kawamoto, Z. M. Fadlullah, H. Nishiyama, and N. Kato, ``A survey on network methodologies for real-time analytics of massive IoT data and open research issues,'' \emph{IEEE Commun. Survey Tuts.}, vol. 19, no. 3, pp. 1457-1477, 3rd Quart., 2017.









%\bibitem{LTE_R13} 3GPP, ``Release 13: Service Requirements Maintenance for Machine-Type Communications (MTC)/,'' Mar. 2016; http://www.3gpp.org.
%
%\bibitem{LTE_R14} 3GPP, ``Release 14: Further enhancements of MTC and NB-IoT,'' Jun. 2017; http://www.3gpp.org.





%\bibitem{LTE_R14} 3GPP, ``TR 21.914 Study on Provision of Low-Cost MTC UE Based on LTE,
%V 0.9.0,'' 2013
% 3GPP RP-161901, "Revised Work Item Proposal: Enhancements
%of NB-IoT,?Sept. 2016


\end{thebibliography}

\end{document}